
\magnification 1000
\baselineskip=12 pt
\font\note=cmr10 at 10 true pt


\def\limz{\buildrel _{z\to 0} \over \longrightarrow}
\def\Klpp{K_L\rightarrow \pi^+\pi^-}
\def\Kspp{K_S\rightarrow \pi^+\pi^-}

\def\Kopp{K^\pm\rightarrow \pi^\pm\pi^0}
\def\Kcppg{K_{L,S}\rightarrow \pi^+\pi^-\gamma}
\def\Knppg{K_{L,S}\rightarrow \pi^0\pi^0\gamma}
\def\Klppg{K_L\rightarrow \pi^+\pi^-\gamma}
\def\Ksppg{K_S\rightarrow \pi^+\pi^-\gamma}

\def\Koppg{K^\pm\rightarrow \pi^\pm\pi^0\gamma}
\def\Klnppg{K_{L}\rightarrow \pi^0\pi^0\gamma}
\def\Ksnppg{K_{S}\rightarrow \pi^0\pi^0\gamma}

\def\vareps{{\widetilde \epsilon}}

\def\Fp{F_{\pi}}
\def\Fz{F_0}
\def\Fk{F_K}

\def\Kpp{K\rightarrow \pi\pi}
\def\Kppg{K\rightarrow \pi\pi\gamma}

\def\DS{{|\Delta S|=1}}
\def\da{^{\dagger}}
\def\dmu{\partial_{\mu}}

\def\dmuu{\partial^{\mu}}
\def\dnuu{\partial^{\nu}}
\def\Amu{A_{\mu}}
\def\Dmu{D_{\mu}}
\def\Dmuu{D^{\mu}}
\def\fmunup{ f^{\mu\nu}_+}
\def\fmunum{ f^{\mu\nu}_-}
\def\lmu{ l_{\mu}}
\def\rmu{ r_{\mu}}
\def\DDrho{\bigtriangledown_{\rho}}
\def\Ed{{E_2}}
\def\Eu{{E_1}}
\def\Real{{\Re e}}


\def\Sol{1}
\def\Report{2}
\def\ENPa{3}
\def\ENPb{4}
\def\Low{5}
\def\WG{6}
\def\GLa{7}
\def\Ramberg{8}
\def\Seghal{9}
\def\EPRc{10}
\def\DMS{11}
\def\condon{12}
\def\EGPR{13}
\def\BEG{14}
\def\Eckerb{15}
\def\Dib{16}
\def\Seghalc{17}
\def\data{18}
\def\ochs{19}
\def\GM{20}
\def\Rambergb{21}
\def\Esposito{22}
\def\KMWnucl{23}
\def\Wolf1{24}
\def\Taureg{25}
\def\abrams{26}
\def\Funck{27}
\def\EPRa{28}
\def\Good{29}


\def\today{June 1994}
\vskip 1cm
\today
\vskip 1cm
\null
\line{\hfill INFNNA-IV-20 6/94}
\line{\hfill ROME-1030 6/94}
\null
\vskip 2.5 true cm
\centerline{{\bf  $\Kppg$ decays:
a search for novel couplings in kaon decays}
\footnote{$ ^*$}{\note
Work supported in part by the Human Capital and Mobility Program, EEC
Contract N. CHRX-CT920026.}}
\vskip 2.0 true cm
\centerline{Giancarlo D'Ambrosio\footnote{ $^1$}{\note
E-mail: dambrosio@vaxna1.na.infn.it}}
\medskip
\centerline{\it Istituto Nazionale di Fisica Nucleare, Sezione di Napoli,
I-80125 Italy}
\centerline{\it Dipartimento di Scienze Fisiche, University of Naples,
I-80125 Italy}
\bigskip
\centerline{and}
\bigskip
\centerline{Gino Isidori\footnote{ $^2$}{\note
E-mail: isidori@vaxrom.roma1.infn.it}}
\medskip
\centerline{\it Istituto Nazionale di Fisica Nucleare, Sezione di Roma
 I-00185 Italy}
\centerline{\it Dipartimento di Fisica, University of Rome, ``La Sapienza",
I-00185  Italy}
\vskip 1cm
\centerline{\bf{Abstract}}
\bigskip
\midinsert\narrower\narrower
\noindent
We analyze $\Kppg$  decays in the framework
of Chiral Perturbation Theory.
We  study the different Dalitz plot distributions, trying to find
regions where  o($p^6$) contributions could be more easily detected.
To fulfill this program
we compute all
the the  o($p^4$) loop and counterterm contributions, finding a
substantial agreement with the existing calculations and adding some
small missing terms in $\Ksppg$.
\endinsert
\vfill
\eject


\vskip 1. true cm
\centerline{\bf{1. Introduction}}
\vskip .8 true cm

The amplitudes for $\Kppg$
decays can be generally decomposed as
the sum of two terms: the internal bremsstrahlung $(A_{IB})$ and
the direct emission $(A_{DE})$ [\Sol-\ENPb].
The first one, which represents only the  bremsstrahlung contribution
of the external charged particles,  is completely predicted
by QED in terms of the  $\Kpp$ amplitude  [\Low] .
The second one, which is obtained by
subtracting $A_{IB}$ from the total amplitude,
depends on the direct $\Kppg$ coupling and  furnishes a test for mesonic
interaction models.

The pole for the photon energy going to zero
would tend to enhance $A_{IB}$ compared to $A_{DE}$. Nevertheless
in $\Klppg$ and $\Koppg$  decays the inner bremsstrahlung
is respectively suppressed by
the $CP$ conservation and the $\Delta I=1/2$ rule. As a result
in these channels it is easier to study the direct emission, testing
low energy interaction models.

Chiral Perturbation Theory (ChPT) is supposed to be
a reliable framework to study $\Kppg$ decays [\WG, \GLa].
In this effective quantum field theory, based on symmetry principles,
the physical amplitudes are expanded in powers of meson
masses and momenta. At the lowest non-trivial order ($o(p^2)$), only
the inner bremsstrahlung  appears, while the direct emission,
which requires more derivatives, starts at $o(p^4)$.
Experiments  seem to indicate
that in $\Klppg$, and possibly in $\Koppg$, not only  $o(p^4)$ but also
$o(p^6)$ magnetic contributions might be important [\Ramberg, \ENPb].
More in general, we can say that the question of the size of
$o(p^6)$ contribution in radiative non-leptonic kaon decays is very
interesting, as it as been emphasized in the case of
$K_L\rightarrow \pi^0\gamma\gamma$ [\Seghal, \EPRc].

In this paper we have analyzed  $\Kppg$ decays,
looking for kinematical regions where order $p^6$ electric contributions
could be relevant and possibly detected. To this aim we have computed
all the one loop amplitudes and the corresponding $o(p^4)$
counterterms contributions. We substantially agree with the existing
calculation for $\Klppg$, $\Koppg$ [\ENPb] and
$\Ksppg$ [\DMS]. In addition we have calculated the small missing
terms due $\pi K$ and  $K\eta$ loops in $\Ksppg$.
Using these results we find some kinematical region,
in particular for $\Klppg$ and possibly for $\Koppg$ decays,
where order $p^6$ electric contributions could be relevant.

The paper is organized as follows: in section 2 we give a short
discussion on kinematics, Low theorem and ChPT. In section 3, 4, 5
and 6 we analyze respectively $\Klppg$, $\Ksppg$, $\Koppg$ and
$\Knppg$ decays. Section 7 contains some concluding remarks and finally in
the appendices we discuss the loop functions and some variable transformations.

\vskip 1. true cm
\centerline{\bf{2. Kinematics, Low theorem and ChPT}}
\vskip .8 true cm

Due to Lorentz and gauge invariance we can define two
invariant amplitudes for the processes $K(p_{_K})\rightarrow\pi_1(p_1)
\pi_2(p_2)\gamma(q,\vareps)$:  the electric and the  magnetic one.
Using the dimensionless amplitudes $E$ and $M$, defined in Ref.[\ENPb],
we can write:
$$
A(\Kppg)=\vareps^\mu(q)\left[E(z_i)(p_1qp_{2\mu}
-p_2qp_{1\mu})+M(z_i)\epsilon_{\mu\nu\rho\sigma}p_1^\nu p_2^\rho q^\sigma
\right] /m_K^3,   \eqno(1)
$$
with
$$
z_i={p_i q \over m_K^2}, \qquad\qquad
z_3={p_{_K} q \over m_K^2}, \qquad\qquad  z_3=z_1+z_2.
\eqno(2)
$$
Summing over photon helicities there  is no interference among electric and
magnetic terms:
$$
{\partial^2\Gamma\over{\partial z_1\partial z_2}}=
{m_K\over{(4\pi)^3}} (|(E(z_i)|^2+|M(z_i)|^2)\left[
z_1z_2(1-2z_3-r_1^2 - r_2^2)-r_1^2z_2^2 -r_2^2z_1^2\right].
\eqno(3)
$$

As we have already said in the introduction, generally  one prefers also
to decompose the total amplitude as a sum of inner bremsstrahlung  and direct
emission:
$$
A(K \rightarrow \pi_1\pi_2\gamma) =A_{IB}+A_{DE}.\eqno(4)
$$
$A_{IB}$, which in the classic limit would correspond to the charged particle
radiation, in QED is completely predicted by the Low theorem [\Low],
which relates radiative and non-radiative amplitudes
in the limit of photon energy going to zero. For $\Kppg$ transitions the
theorem  reads:
$$
A_{IB}(\Kppg) =e\left({\vareps p_b \over qp_b }-{\vareps p_a \over qp_a }
\right) A(\Kpp),\eqno(5)
$$
where $(p_a,p_b)\equiv (p_+,p_-)$ for the neutral kaon decays
and $(p_a,p_b)\equiv (p_+,p_K)$ or $(p_a,p_b)\equiv (p_K,p_-)$ for the charged
 ones.
Thus in the $E(z_i)$ amplitude the two contributions  $A_{IB}$ and
$A_{DE}^{Electric}$ can interfere, differently from the $M(z_i)$
amplitude, where only a direct emission contribution appears:
$$
|A(K \rightarrow \pi_1\pi_2\gamma)|^2 =|A_{IB}|^2+2\cdot{\cal R }e
\{ A_{IB}^*A_{DE}^{Electric}\}+|A_{DE}^{Electric}|^2+|A_{DE}^{Magnetic}|^2.
\eqno(6)
$$
By opportune kinematic integration, eq.(3) can be transformed in a photon
energy distribution (see appendix B)
. In the limit of the photon energy ($E_\gamma^*$ in
the kaon rest frame)  going to zero one can write:
$$
{d\Gamma\over{ dE_\gamma^*}}
={\alpha\over E_\gamma^*}+\beta+\gamma \cdot E_\gamma^*+...\eqno(7)
$$
The content of Low theorem is a prediction for $\alpha$ and $\beta$,
from pure QED, in terms of the physical amplitude  $A(\Kpp)$.
\medskip
We will now proceed analyzing $\Kppg$ decays in the framework of ChPT.
In this effective quantum field theory the octet of the pseudoscalar
fields\footnote{$\da$}{\note The chosen phase convention does not satisfy the
Condon-Shortley-De Swart one [\condon].},  $$
\phi=\left[\matrix{& \pi^0/\sqrt{2} +\eta/\sqrt{6} & \pi^+ &
K^+ \cr & \pi^- & -\pi^0/\sqrt{2} +\eta/\sqrt{6} & K^0 \cr
& K^- & {\overline K}^0 & -\eta \sqrt{2/3}
\cr}\right], \eqno(8)
$$
is identified with the octet of pseudo Goldstone bosons coming from
the spontaneous symmetry breaking of $G=SU(3)_L\times SU(3)_R$ into
$H=SU(3)_{L+R}$. The lagrangian, expressed in terms of $\phi$,
is expanded in powers of meson masses and momenta, and is dictated only by
the transformation law of the interactions.

At the lowest order $p^2$, the strong part of the lagrangian is
completely fixed and is  given by:
$$
{\cal L}^{(2)}_S = {\Fz^2 \over 4}\left\lbrace {\rm tr}(
\Dmu U\da \Dmuu U)+ {\rm tr}(\chi U\da+U\chi\da) \right\rbrace,\eqno(9)
$$
where $U=e^{i\sqrt{2}\phi/\Fz}$
is a unitary $3\times 3$ matrix which transforms linearly under $G$ and
$\chi$  is proportional to a scalar field, transforming like $U$, which must
acquire a non-vanishing expectation value in order to reproduce the correct
values of the meson masses [\GLa]. At this order we can assume
$\Fz=\Fp=\Fk\simeq 93{\rm MeV}$. The covariant derivative $\Dmu U$ is
given by:
$$
\Dmu U=\dmu U-i\rmu U +iU\lmu, \eqno{(10)}
$$
where $\lmu$ and $\rmu$ are the external gauge fields of $SU(3)_L\times
SU(3)_R$. The electromagnetic field can be introduced setting
$\lmu=\rmu=eQ\Amu$, where $Q={\rm diag}(2/3, -1/3, -1/3)$.
\medskip
The $\DS$ weak lagrangian transforms under $G$ as an $(8_L,1_R)$ or
a $(27_L,1_R)$. At the lowest order $p^2$ can be written as:
$$
\eqalign{
{\cal L}_\DS^{(2)} &=G_8 \Fp^4 {\rm tr}\left(\lambda \Dmu U\da\Dmuu U\right)\cr
&+G_{27} \Fp^4 \left[ (U\da\Dmu U)_{23}(\Dmu U\da U)_{11}+ {2\over 3}
(U\da\Dmu U)_{21}(\Dmu U\da U)_{13} \right] +{\rm h.c.},} \eqno(11)
$$
where $\lambda=(\lambda_6-i\lambda_7)/2$ and $\lambda_{6,7}$ are the
usual Gell-Mann matrices. Experimentally, from $\Kpp$ decays, the coupling
constant $G_8$  and $G_{27}$ are fixed to be:
$$
|G_8|\simeq 9\cdot 10^{-6}\ GeV^{-2}, \qquad \qquad {G_{27}
\over G_8}\simeq{1\over 18}.\eqno(12)
$$
At order $p^2$, in $\Kppg$ decays, since there are not enough powers of
momenta, only an inner bremsstrahlung amplitude will appear.  In
agreement with the Low theorem we found:
$$
A_{IB}(K \rightarrow \pi\pi\gamma)^{o(p^2)} =
e\left({\vareps p_b \over qp_b }-{\vareps p_a \over qp_a }
\right) A(K\rightarrow \pi\pi)^{o(p^2)}.\eqno(13)
$$
Actually we will take eq.(13) as a definition of the inner bremsstrahlung
amplitudes, meaning that this relation must hold order by order in ChPT.
We recall also the $o(p^2)$ results for the CP conserving $\Kpp$ amplitudes:
$$
\eqalign{
A(K_{S} \rightarrow \pi^+\pi^-)^{o(p^2)}=&
+2( G_8+ {2\over 3} G_{27})\Fp(m_K^2-m_\pi^2),   \cr
A(K^+ \rightarrow \pi^+\pi^0)^{o(p^2)}=&
+{5\over 3} G_{27}\Fp(m_K^2-m_\pi^2). \cr} \eqno(14)
$$

\medskip
ChPT is not a renormalizable theory but the requirement of unitarity implies
that meson loops have to be considered. The loop contributions, which start
at order $p^4$, introduce divergences which must be re-absorbed,
order by order, by corresponding counterterms. While the
physical amplitudes (loops + counterterms) are scale independent, the
coefficients of the counterterms depend on the renormalization
scale of the loops.
Gasser and Leutwylwer [\GLa] have classified the $p^4$ operators for the
strong and electromagnetic lagrangian and have determined their coefficients
by comparison with the experiments.
The authors of Ref.[\EGPR] have shown that these coefficients are well
reproduced
by vector and axial vector meson exchange choosing the renormalization scale
$\mu$ around the $\rho$-mass.

The effective Lagrangian at next to leading order, both in the strong and weak
sector, has anomalous and non-anomalous operators [\BEG]. At this order the
anomalous operators contribute only to the magnetic amplitudes $M(z_i)$,
which have been extensively studied in Ref.[\ENPa, \ENPb] and will be not
re-analyzed here. The non-anomalous octet part of the $p^4$ $\DS$ Lagrangian,
relevant for $\Kppg$ decays, can be written as:
$$
{\cal L}^{(4)}_\DS = G_8 \Fp^2 [N_{14} W_{14}+N_{15} W_{15}+N_{16} W_{16}+
N_{17} W_{17}] +{\rm h.c.}, \eqno(15)
$$
where
$$
\eqalign{
W_{14} &= i{\rm tr} \Big(\lambda u\da\Big\lbrace
\fmunup , u_\mu u_\nu \Big\rbrace u \Big), \cr
W_{15} &= i {\rm tr} \Big(\lambda u\da u_\mu\fmunup
u_\nu  u \Big),  \cr
W_{16} &= i {\rm tr} \Big(\lambda u\da\Big\lbrace
\fmunum , u_\mu u_\nu \Big\rbrace u \Big),  \cr
W_{17} &=i {\rm tr} \Big(\lambda u\da u_\mu\fmunum
u_\nu  u \Big), \cr} \eqno(16)
$$

$$\lambda=(\lambda_6-i\lambda_7)/2,\quad
u^2=U, \qquad u_\mu=i u\da \Dmu U u\da =u_\mu\da, \qquad
f^{\mu\nu}_\pm = u F_L^{\mu\nu} u\da \pm u\da F_R^{\mu\nu} u \eqno(17)
$$
and
$$
F_L^{\mu\nu}  = \dmuu l^\nu- \dnuu l^\mu +i[l^\mu,l^\nu], \qquad\quad
F_R^{\mu\nu}  = \dmuu r^\nu -\dnuu r^\mu +i[r^\mu,r^\nu]. \eqno{(18)}
$$
In the electric transitions of $\Kppg$ decays
only the following counterterm combination appears:
$$N_{14}-N_{15}-N_{16}-N_{17}. \eqno{(19)}$$
Chiral symmetry alone does not predict the values of the constants $N_i$,
these however can be estimated using various Vector Meson Dominance (VMD)
models. Following Ref.[\Eckerb], the combination in eq.(19) can be
expressed in terms of the parameter $k_f$:
$$
N_{\Eu}^{(4)}=(4\pi)^2\left[N_{14}-N_{15}-N_{16}-N_{17}\right]
=-k_f { 8\pi^2 F^2_{\pi} \over M_V^2 }\sim o(1). \eqno{(20)}
$$
\medskip
At $o(p^6)$ obviously the number of E-type counterterms increases,
normalizing ${\cal L}^{(6)}$ as ${\cal L}^{(2)}$ and
${\cal L}^{(4)}$ we can write in general:
$$
{\cal L}^{(6)}_{\DS}=G_8  \sum_iN^{(6)}_i W^{(6)}_i +{\rm h.c.} \eqno(21)
$$
We will not try to classify all the $W^{(6)}_i$ but we will consider
only two typical operators:
$$
\eqalignno{
W_{\Eu}^{(6)}=&{i\over(4\pi)^4}{\rm tr}\left(u\lambda u\da
\left( u\chi\da u u_\mu \fmunum u_\nu +  u_\mu \fmunum u_\nu u\da\chi
u\da \right) \right), &(22) \cr
W_{\Ed}^{(6)}=&{1\over(4\pi)^4}{\rm tr}\left( u\lambda u\da
\left\lbrace \fmunup,\left[ u^\rho, \DDrho [u_\mu,u_\nu] \right]
\right\rbrace  \right), &(23) \cr }
$$
($\DDrho$ is the "covariant'' derivative defined in ref.[\EGPR],
which for our purposes acts like $\partial_\rho$).
The first one, similar to the one
considered in Ref.[\Dib], generates the same kinematical
dependence of the $o(p^4)$ operators, i.e. give a contribution to the
electric amplitudes of $\Kppg$ decays proportional to the one of the $o(p^4)$
counterterms. The second one, with more derivatives, generates a
substantially different kinematical dependence. The size of the coefficients
of these operator is an open question. With the normalization in
eqs.(22)-(23), naive chiral power counting would suggest $N_{\Eu}^{(6)}
\sim N_{\Ed}^{(6)} \sim o(1)$ nevertheless, for different arguments,
sensibly larger estimates for $N_{\Eu}^{(6)}$ and $N_{\Ed}^{(6)}$ were made
respectively in Ref.[\Dib] and [\Seghalc]. Waiting for further theoretical
insight, we will not commit ourself to any particular model to predict the
size of these coefficients, but we will investigate the
possibility to bound or even to measure them.

\vskip 1. true cm
\centerline{\bf 3.   $\Klppg$ }
\vskip .8 true cm
In this decay the electric and magnetic contribution of eq.(3)
are competing and they have been both measured [\Ramberg].
The inner bremsstrahlung  violates CP
through the $\Klpp$ amplitude and, neglecting the $\Delta I=3/2$ suppressed
terms, can be written as:
$$
E_{IB}(z_i)=+{\epsilon e A(\Kspp) \over m_K (z_+ z_-)},
\qquad \qquad (p_1 =p_+,\ p_2=p_-) , \eqno(24)
$$
which at the lowest order in ChPT is:
$$E_{IB}^{o(p^2)}(z_i)={2e\epsilon  G_8\Fp(m_K^2- m_\pi^2) \over m_K (z_+
z_-)} . \eqno(25)
$$
To decrease the theoretical uncertainties it is wiser to use in
eq.(24) [\DMS] the experimental value for $| A(\Kspp) |$ [\data]
and the phases
according the most recent analysis based on very general
assumptions like analyticity,  unitarity and crossing symmetries  [\ochs, \GM].
 Of course one has to use  the phase convention of eq.(11).

The direct emission contribution can be generally decomposed in a
multipole expansion [\Report]:
$$
\eqalign{
E_{DE} =& E_1(z_3)+E_2(z_3)\cdot (z_+ - z_-) +... \cr
     M =& M_1(z_3)+M_2(z_3)\cdot (z_+ - z_-) +... \cr  }\eqno(26)
$$
We remind that even electric and odd magnetic multipoles are CP even,
while the others are CP odd. The multipole expansion is particularly
justified in these decays by the smallness of the phase space, which implies
$|z_+ - z_-|<0.2$. Furthermore, local contribution to higher multipoles are
suppressed by chiral power counting.
So far only the inner bremsstrahlung and the
dipole magnetic contribution
have been observed. The results for the branching ratios are [\Ramberg]:
$$\eqalign{
BR(\Klppg)_{IB,E_\gamma^*>20 \ MeV}=(1.32\pm.05)10^{-5},  &\cr
BR(\Klppg)_{DE,E_\gamma^*> 20\ MeV}=(2.95\pm.11)10^{-5}. &\cr}
\eqno(27)$$
For the IB  both the spectrum and the rate are in agreement with eq.(24).
The dipole magnetic contribution can be written in first approximation as
$$
M_1(z_3)\simeq M_1^0\cdot(1+\hat{c}z_3), \eqno(28)
$$
where $M_1^0$ and $\hat{c}$ are fixed by the rate and the spectrum
respectively.
The large experimental vale for the slope:
$\hat{c} ^{exp}=(-1.7\pm 0.5)$ [\Rambergb], which vanishes at $o(p^4)$ in ChPT,
carries an
important dynamical informations. It is not only a pure $o(p^6)$
contribution, but it is also substantially larger of a typical vector
meson exchange effect [\ENPb],
showing that, at least in the weak anomalous sector, local $p^6$ contributions
are sizable and can play a role in discriminating different models.

Regarding the electric multipoles,
at $o(p^4)$ there are no local CP conserving counterterms and consequently
the loop contributions are finite [\Esposito, \KMWnucl]. We have computed
these amplitudes confirming the earlier results of Ref.[\ENPb].
The potential contributions are given by $\pi\pi$, $\pi K$, $K \eta$  and $KK$
intermediate states (see fig.1). Nevertheless, due to CP conservation,
only the small $\pi K$ and $K \eta$  loops are non-vanishing.
Their contributions can be parametrized using the functions
$h_{\pi K}$ and $h_{K\eta}$ defined in the appendix A:
$$
E^{(4)}_{loop}(z_i)={ e G_8 m_K (m_K^2 -m_\pi^2) \over 8\pi^2\Fp} \left[
 h_{\pi K}( z_-)+h_{K\eta}( z_-) - h_{\pi K}( z_+)-h_{K\eta}( z_+) \right].
\eqno(29)
$$
The functions $h_{ij}$
depend only on one kinematical variable. Considering the Taylor expansion
of these functions ($h_{ij}(z_i)=\sum_n h_{ij}^n z_i^n$), with the
coefficients reported table 1, we found:
$$
E^{(4)}_{loop}(z_i)\simeq {e G_8 m_K (m_K^2 -m_\pi^2) \over 8\pi^2\Fp}
\left[\ 0.05\ (z_+-z_-) \ \right].
\eqno(30)
$$
Consistently with the CP conservation only an
$\Ed$ term survives but it is suppressed by several factors compared to
$E_{IB}$
in eq.(25): i) the angular momentum barrier, ii) the absence of the photon
energy pole, iii) the chiral suppression factor $({m_K\over {4\pi\Fp}})^2$ and
iv) the small coefficient 0.05 generated by the chiral loops.

Other $\Ed$ contribution can arise at $o(p^6)$.
The operator $W_{\Eu}^{(6)}$ is forbidden by CP conservation,
like the $o(p^4)$ counterterms,
on the other hand the operator $W_{\Ed}^{(6)}$ in eq.(23) gives:
$$
E^{(6)}_{\Ed}(z_i)=+{ e G_8 m_K^5 \over 48\pi^4\Fp^3} N^{(6)}_{\Ed}
(z_+-z_-). \eqno(31 )
$$
Since $N^{(6)}_{\Ed} \sim o(1)$ the $\Ed$ multipole
should be dominated by the local $o(p^6)$ counterterms instead of the $o(p^4)$
loop contributions. This is one of the main message  of this paper. At higher
order chiral loops will generate a small rescattering phase $\delta_{\Ed}$ to
the term in eq.(31).

Now the question is whether the $\Ed$ multipole can be detected. The
electric contribution to eq.(3) is
$$
\eqalign{
|E_{IB}+E_{DE}|^2 & \simeq \left[ |E_{IB}|^2+ {2\Real \left( E^*_{IB}E_{\Ed}
\right)}\right] \cr  & = {{|\epsilon e A(\Kspp)|^2 }\over m_K^2
z_+^2z_-^2}  \left[1+C_{int}^L\ z_+z_-(z_+- z_-)\right],} \eqno(32)
$$
where
$$
C_{int}^L \simeq
+{  m_K^4 \cos( \phi_\epsilon + \delta^0_0-\delta_{\Ed})
 \over 48 |\epsilon| \pi^4 \Fp^4 } N^{(6)}_{\Ed},\eqno(33)
$$
$ \delta^0_0\simeq(39\pm 5)^\circ$ [\ochs] is the the  $(\pi^+ \pi^-)_{I=0}$
phase shift,  $\phi_\epsilon \simeq (43.67\pm 0.14)^\circ$[\Wolf1] is the phase
of $\epsilon$ and $\delta_{\Ed}$ is expected to be small.
Unfortunately the argument of the cosine in eq.(33) tends to be about
80$^\circ$, smearing somewhat the effect. Nevertheless we can try to study the
interference term  $C_{int}^L$, and thus $N^{(6)}_{\Ed}$, measuring the
(indirect) CP violating asymmetry  under pion exchange, i.e. the observable:
$$
\Delta^L = { N(\Klppg ;\ \cos(\theta)>0 ) -  N(\Klppg ;\ \cos(\theta)<0 )
\over
 N(\Klppg ;\ \cos(\theta)>0 ) +  N(\Klppg ;\ \cos(\theta)<0 ) }, \eqno(34)
$$
where $\theta$ is the angle between $\gamma$ and $\pi^+$ momenta in the
di-pion rest frame (see appendix B). For $E_\gamma^* > 20$ MeV
$$
\vert \Delta_{(E_\gamma^*>20 \ MeV)}^L \vert= 1.2\cdot 10^{-4}\vert C_{int}^L
\vert
\sim  10^{-3} \vert N^{(6)}_{\Ed} \vert, \eqno(35)
$$
thus at least $10^{11}\ K_L$  are necessary in order to observe a non vanishing
effect.

We finally note that $\Delta^L$ could receive a contribution also from the
interference between $\Ed$ and a direct CP violating multipole $\Eu$.
Nevertheless this contribution it is not enhanced by the interference with the
inner bremsstrahlung and therefore should be suppressed respect to the previous
one.

\vskip 1. true cm
\centerline{\bf 4.  $\Ksppg$ }
\vskip .8 true cm

In this channel we will consider only the electric amplitudes
since the dipole magnetic contribution is CP violating while higher multipoles
are phase space suppressed and do not interfere with electric amplitudes for
unpolarized photons. The  CP conserving bremsstrahlung term
$$
E_{IB}(z_i)=+{e A(\Kspp) \over m_K (z_+ z_-)},
\qquad \qquad (p_1 =p_+,\ p_2=p_-),  \eqno(36)
$$
dominates giving a contribution to the branching ratio
$$
BR(\Ksppg)_{IB,E_\gamma^*>20 \ MeV} \simeq 4.80 \cdot
10^{-3}\eqno(37)
$$
and a Dalitz Plot distribution, reported for instance in Ref.[\Taureg,
\Report],
which agree very nicely with the data [\Ramberg].

At the lowest order in ChPT only the IB is present, the electric dipole
component appear at the next order. The two gauge invariant contributions
$E_{IB}$
and $E_{DE}$ are separated according to the ChPT definition of IB (eq.(13)).
At order o($p^4$) there are loops and counterterms.
Since the counterterm contribution to $A(\Ksppg)_{DE}$
$$
E^{(4)}_{CT}(z_i)=+{4 e G_8 m_K^3 \over\Fp} \left[N_{14}-N_{15}-N_{16}-N_{17}
\right] = +{e G_8 m_K^3  \over 4\pi^2\Fp} N_{\Eu}^{(4)} \eqno(38)
$$
is scale independent [\Esposito, \KMWnucl], the direct emission
loop contribution has to be finite [\DMS]. As in the $K_L$ case the  possible
intermediate states appearing in the meson loop are
$\pi\pi$, $\pi K$, $K \eta$ and $K K$ (see fig.1). In Ref.[\DMS]
only the $\pi\pi$ loop was computed, since this is the only one which
has an absorptive part. Interestingly a large dispersive
contribution was found, even larger than the absorptive one in the allowed
kinematical regions. This of course urged a calculation of the other dispersive
contributions and a  confirmation of the previous $\pi \pi$ loop calculation.
Analogously to eq.(29), our result for the loop contribution
can be written as:
$$
E^{(4)}_{loop}(z_i)=-{e G_8 m_K (m_K^2 -m_\pi^2) \over 8\pi^2\Fp} \left[
4h_{\pi\pi}(-z_3)+h_{\pi K}( z_+)+h_{K\eta}( z_+) + (z_+ \leftrightarrow z_-)
\right], \eqno(39)
$$
where $h_{\pi \pi}$,  $h_{\pi K}$ and $h_{K\eta}$ are defined in the appendix
A (the ${KK}$ contribution  vanishes in the limit $m_{K^+}=m_{K^0}$).
Regarding the $ \pi\pi$ loop we confirm the interesting result that the
dispersive contribution is larger than the absorptive one, however we disagree
on the analytic structure. We find indeed
$$
\eqalign{
E^{(4)}_{\pi\pi}(& E_\gamma^*)=-{e G_8 m_K \over 8\pi^2\Fp E_\gamma^{*2}}
\left\{ (m_K^2-m_\pi^2)(m_K^2-2E_\gamma^*m_K) \left[\beta \ln\left({{1
+\beta}\over {\beta-1}}\right)-  \beta_0 \ln\left({{1 +\beta_0}\over
{\beta_0-1}}\right)\right]\right. \cr  &
\left.-E_\gamma^*m_K(2m_\pi^2-2m_K^2)+m_\pi^2(m_K^2-m_\pi^2)
\left[\ln^2\left({{1+\beta_0}\over{\beta_0-1}}\right)-
\ln^2\left({{1+\beta}\over{\beta-1}}\right)\right]\right\},  }
\eqno(40)
$$
while the authors of Ref.[\DMS] for the polynomial term had
$(3m_\pi^2-2m_K^2)$. Though the difference is numerically insignificant
has some deep meaning: i) the full amplitude is proportional to
$(m_K^2-m_\pi^2)$ , i.e. to the weak vertex of fig.1 with the pions on-shell,
ii) the  $E^{(4)}_{loop}(E^*_\gamma)$ amplitude vanishes in the limit
$E^*_\gamma\to
0$, as required by Low theorem\footnote{$\da$}{\note
The author of Ref.[\DMS] have recently checked their
calculation confirming our result.}.
Of course with the help of these considerations one understands
better
why the $KK$ loop  contribution is zero. Furthermore it is also confirmed  by
our calculation that the $\pi\pi$ loop contribution is dominant
as  suggested in  Ref.[\DMS].

Another interesting point of our calculation is the kinematical dependence of
the loop contributions. Using the Taylor expansion of the function $h_{ij}$ in
eq.(39) we obtain:
$$
E^{(4)}_{loop}(z_i)\simeq -{e G_8 m_K (m_K^2 -m_\pi^2) \over 4\pi^2\Fp} \left[
\ 1.3\ +\  1.1 z_3-i(0.6+0.9 z_3)\   \right]. \eqno(41)
$$
Since the kinematical dependence of eq.(41) is  mild, in principle it could
be possible to distinguish particular $o(p^6)$ contributions, like
the one of the operator $W^{(6)}_\Ed$:
$$
E^{(6)}_{\Ed}(z_i)= -{ e G_8 m_K^3 \over 24\pi^4\Fp^3} N^{(6)}_{\Ed}
\left[ m_K^2- m_\pi^2 -{3\over 2}m_K^2 z_3 \right]. \eqno(42)
$$
On the other hand it is impossible to distinguish among the flat contributions
of the $o(p^4)$ counterterms and the operator $W^{(6)}_\Eu$:
$$
E^{(6)}_{\Eu}(z_i)=-{ e G_8 m_K^5 \over 32 \pi^4\Fp^3} N^{(6)}_{\Eu}.
\eqno(43)  $$
Of course all these contributions give a very  small interferencial branching
ratio compared to the IB in eq.(37). As explained in the discussion regarding
the bremsstrahlung of $\Klppg$, for  $A(\Kspp)$ we will use the experimental
value of $|A(\Kspp)|$ and  $ \delta^0_0\simeq(39\pm 5)^\circ$   [\GM, \ochs].
Neglecting the $p^6$ contribution, using the central value of $ \delta^0_0$
and varying  $N_{\Eu}^{(4)}$ we obtain the results reported in table 2.

Differently from the $K_L$ case, due to the large IB, in $\Klppg$ decays it is
practically impossible to measure a CP violating interference between an even
and an odd multipole.

\vskip 1. true cm
\centerline{\bf 5.  $\Koppg$ }
\vskip .8 true cm

Since the initial and the final states of $\Koppg$ decays are not CP
eigenstates,
in
these channels the inner bremsstrahlung amplitude together with
the lowest electric and magnetic transitions are present, also in the limit
of CP conservation. Actually $E_{IB}$ is suppressed by the ${\Delta I=1/2}$
rule:
$$
E_{IB}(z_i)={\pm e A(\Kopp) \over m_K (z_\pm z_3)},
\qquad \qquad (p_1 =p_\pm,\ p_2=p_0).\eqno(44)
$$
Experimentalists generally choose one kinematical variable as the kinetic
energy of the charged pion ($T_c^*$ in the
CMS), which is not affected by the problem of identifying
the $\pi^0$ photons. In the range  $55\ $MeV$\le T_c^*\le 90\ $MeV
the theoretical prediction for internal bremsstrahlung branching ratio is:
$$
BR(\Koppg)_{IB,\ 55 MeV\le T_c^*\le 90 MeV }=2.61\cdot 10^{-4}. \eqno(45)
$$
After the subtraction of this contribution from the experimental branching
ratio there is a clear evidence of a direct emission component [\data]:
$$
BR(\Koppg)_{DE,\ 55 MeV\le T_c^*\le 90 MeV }
=(1.8\pm 0.4)\cdot 10^{-5}. \eqno(46)
$$
It is only one order of magnitude less than the IB in eq.(45), instead
of the typical $10^{-2} \sim 10^{-3}$ suppression factor of the interferencial
contribution, because both  the electric and magnetic amplitudes
are not suppressed by the ${\Delta I=1/2}$ rule.

{}From the analysis of the Dalitz Plot distribution, the DE component seems
more likely due to a magnetic transition [\abrams], but the present data are
not
conclusive about this point. The theoretical discussions of the magnetic
amplitudes has been done for instance in Ref.[\ENPa, \ENPb]. Here we will
concentrate on the electric ones in the framework of ChPT.
At the lowest order, as usual, there is only the IB contribution:
$$
E_{IB}^{o(p^2)}(z_i)=\pm{5e G_{27} \Fp(m_K^2- m_\pi^2) \over 3m_K
(z_\pm z_3)}.\eqno(47)
$$
At next order there are loops and counterterms. The $o(p^4)$ counterterm
combination which appears is the same finite one of eq.(38):
$$
E^{(4)}_{CT}(z_i)=\mp{e G_8 m_K^3  \over 8\pi^2\Fp} N_{\Eu}^{(4)},\eqno(48)
$$
thus the loop contributions are finite. In the octet limit ($G_{27}=0$) only
the small $\pi K$ and $K \eta$ loops are non-vanishing. We have computed  them
finding a little discrepancy with the earlier
computation of Ref.[\ENPb]. Our result is:
$$
\eqalign{
E^{(4)}_{loop}(z_i)= & \mp{e G_8 m_K (m_K^2 -m_\pi^2) \over 8\pi^2\Fp} \left[
h_{\pi K}( z_+)+h_{K\eta}( z_+) \right]    \cr
 \simeq & \pm{e G_8 m_K (m_K^2 -m_\pi^2) \over 8\pi^2\Fp} \left[\ .16\ +\
.05 z_+\ \right],  }\eqno(49)
$$
weather in Ref.[\ENPb] the $h_{K\eta}$ function was multiplied by a factor
$2/3$\footnote{$\da$}{The author of Ref.[\ENPb] have recently
checked their calculation confirming our result.}.
Due to the smallness of the loop contributions this discrepancy is
numerically irrelevant.

As in the $\Klppg$ case, in $\Koppg$ decays it turns out that the loop
contributions are very small and thus the $o(p^6)$ local operators may be
relevant. However  in this case the $o(p^4)$ counterterms are not vanishing
and should represent the dominant effect. To disentangle the $o(p^6)$
contributions from the $o(p^4)$ ones it is necessary to look for particular
 kinematical dependencies. As in the previous cases the operator $W^{(6)}_\Eu$
gives only a flat contribution which renormalizes the one of the $o(p^4)$
counterterms:
$$
N^{(4)}_{\Eu} \to N^{(4)}_{\Eu}
-{  m_K^2 \over 8 \pi^2\Fp^2} N^{(6)}_{\Eu}.  \eqno(50)
$$
On the other hand the operator $W^{(6)}_\Ed$ gives a non-flat contribution:
$$
E^{(6)}_{\Ed}(z_i)=\mp{ e G_8 m_K^3 \over 24\pi^4\Fp^3} N^{(6)}_{\Ed}
\left[ m_K^2- m_\pi^2 -{3\over 2}m_K^2 z_3 \right], \eqno(51)
$$
which could be detected studying the $T_c^*$ dependence of the Dalitz Plot
distribution (see appendix B).

\vskip 1. true cm
\centerline{\bf 6.  $\Knppg$ }
\vskip .8 true cm

Since no charged particles are involved, $\Knppg$ decays are completely
different from those considered before. The absence of charged particles
implies the absence of the inner bremsstrahlung amplitude and thus a very
small branching ratio. Furthermore these decays are suppressed by Bose
statistics, which requires both electric and magnetic multipoles
to be even. As a consequence none of them has been observed yet.

In the limit of CP conservation $\Klnppg$ is a pure electric transition and
$\Ksnppg$ is a pure magnetic one. We will consider only the former.
As it has been shown in Ref.[\Funck], in $\Klnppg$ not only the $o(p^4)$
local contributions are forbidden, but also the one-loop amplitude is
vanishing.
Therefore this decay can receive contributions  at least of order $p^6$ in ChPT
and could be very useful in order to fix, or even to bound,
these new couplings. Unfortunately it is very difficult to observe,
due to the large background of  $K_L\to 3\pi^0$.

A typical operator which gives a non vanishing contribution to
$\Klnppg$ is $W_\Ed^{(6)}$:
$$
E^{(6)}_{\Ed}(z_i)={ e G_8 m_K^5 \over 48\pi^4\Fp^3} N^{(6)}_{\Ed}
(z_1-z_2). \eqno(52)
$$
Using eq.(52) we can parametrize the branching ratio of $\Klnppg$
in terms of  $N^{(6)}_{\Ed}$:
$$
BR(\Klnppg)\simeq 10^{-9}|N^{(6)}_{\Ed} |^2. \eqno(53)
$$
In Ref.[\Seghalc] the optimistic guess of $BR(\Klnppg)\simeq 10^{-8}$,
together with $BR(\Ksnppg)\simeq 10^{-11}$, was made. This prediction
would suggest $N^{(6)}_{\Ed}\sim 3$, which is certainly bigger than the
naive power counting estimate but cannot be excluded {\it a priori}.

\vskip 1. true cm
\centerline{\bf 7.  Conclusions }
\vskip .8 true cm
The analysis presented here completes the previous $o(p^4)$ calculation
on $\Kppg$ decays and explores the effects of $o(p^6)$ local operators in the
electric transitions.

Regarding the $o(p^4)$ DE loop contributions, we find substantial agreement
with
the existing calculations. In Ref.[\DMS] it was shown that the dispersive
pion loop contribution is larger than the absorptive one. This has urged us
to calculate also the $\pi K$ and $\eta K$ loops, which give only
a dispersive contribution. We find that the naive expectation of Ref.[\DMS],
that these contribution are small, is confirmed. Furthermore our calculation
has  clarified some erroneous considerations
of Ref.[\DMS].

We have also analyzed the contributions to $\Kppg$ decays
of two typical $o(p^6)$ operators (eqs.(22)-(23)), which have been studied
for different purposes in the literature.
Interestingly we find that in $\Klppg$ the
local $o(p^6)$ contributions can be substantially larger than
the $o(p^4)$ ones.

\vskip 1. true cm
\centerline{\bf Acknowledgements}
\vskip .8 true cm

We would like to thank  G. Ecker and H. Neufeld for very useful discussions.

\vfill\eject

\vskip 1. true cm
\centerline {\bf  Appendix A}
\vskip .8 true cm

Following Ref.[\ENPb], in order to write the loop functions $h_{ij}(z)$ we
first introduce the expansion of the three point integrals with $q^2=0$:
$$
\eqalign{
\int {d^4 l \over (2\pi)^4}\ & {l^\mu l^\nu  \over [l^2 - m_i^2]
[(l+q)^2 - m_i^2] [(l-p)^2 - m_j^2] } =  \cr
&\qquad\qquad i g^{\mu\nu} C_{20}( p^2, (p+q)^2,
m_i^2, m_j^2)\ +\ ``p^\mu ,q^\mu\ {\rm terms}". } \eqno(A.1)
$$
With this definition, the finite functions $h_{ij}(z)$ are given by:
$$
h_{ij}(z) = (4\pi)^2 {m_K^2 \over  pq} \left[ C_{20}( p^2,(p+q)^2, m_i^2,m_j^2)
 -  C_{20}( p^2,p^2, m_i^2,m_j^2) \right],  \eqno(A.2)
$$
where $z=pq/m_K^2$ and $p^2=m_\pi^2$ (for $h_{\pi K}$ and $h_{K \eta}$)
or  $p^2=m_K^2$ (for $h_{\pi\pi}$).\par
These functions can be explicitly  written in terms of one
dimensional integrals. Defining:
$$
\eqalign{
f_1(z)&= -{m_i^2 \over z m_K^2} \int_0^1 {dx \over x}\ \ln\left({m_i^2 (1-x)
+xm_j^2 -x(1-x)(p^2+2zm_K^2)   \over   m_i^2 (1-x) +xm_j^2-x(1-x)p^2}
\right), \cr
f_2(z)&= {p^2+m_i^2-m_j^2 +2zm_K^2 \over2 z m_K^2} \int_0^1 dx\ \ln\left(
{m_i^2 (1-x) +xm_j^2 -x(1-x)(p^2+2zm_K^2)   \over   m_i^2 (1-x)
+xm_j^2-x(1-x)p^2} \right), \cr
f_c&= {m_i^2 \over p^2} \int_0^1 dx\ \ln\left({m_i^2 (1-x)
+xm_j^2 -x(1-x) p^2    \over   m_i^2}\right)\ +\  {m_i^2 \over p^2}\cr
&\quad -{m_j^2 \over p^2} \int_0^1 dx\ \ln\left({m_i^2 (1-x)
+xm_j^2 -x(1-x) p^2    \over   m_j^2}\right)\ -\  {m_j^2 \over p^2}\ - 1,\cr}
\eqno(A.3)
$$
we can write:
$$
h_{ij}(z)={1\over 4 z} \left[ f_1(z) + f_2(z) +f_c \right]. \eqno(A.4)
$$

As expected by the  Low theorem, $\left(f_1(z)+f_2(z)\right)\limz -f_c$, so
that $h_{ij}(z)$ has no pole for $z\to 0$ . \par
In the case $m_i=m_j$ the integrals $f_1$, $f_2$ and $f_c$ can be done
explicitly. For the $\pi\pi$ function ($m_i=m_j=m_\pi$, $p^2=m_K^2$) we
found:
$$
\eqalign{
h_{\pi\pi}(z) =&{1 \over 8 z^2}\left\lbrace (1+2z)\left[\beta\ln\left({1+\beta
\over \beta-1}\right) -\beta_0\ln\left({1+\beta_0
\over \beta_0-1}\right)\right] \right. \cr
&\left. \qquad + {m_\pi^2\over m_K^2} \left[\ln^2\left({1+\beta_0 \over
\beta_0-1}
\right)-\ln^2\left({1+\beta \over \beta-1 }\right) \right]
- 2z \right\rbrace,} \eqno(A.5)
$$
where
$$
\beta_0=\sqrt{1-{4m_\pi^2\over m_K^2}}\qquad{\rm and}\qquad
\beta=\sqrt{1-{4m_\pi^2\over m_K^2(1+2z)}}.\eqno(A.6)
$$

It is interesting to no note that the function $h_{\pi\pi}(z)$ is simply
related to the function $H(z)$, defined in Ref.[\EPRa], which appears in $K\to
\gamma\gamma^*$ decays\footnote{$\da$}{We thank G. Ecker for clarifying us this
point}:
$$
h_{\pi\pi}(z)=-H(1+2z). \eqno(A.6)
$$

\vskip 1. true cm
\centerline {\bf  Appendix B}
\vskip .8 true cm

In $\Kcppg$ decays, the variables which can be more easily studied from the
experimental point of view are: i) the photon energy in the kaon rest frame
($E_\gamma^*$), ii)  the angle between $\gamma$ and $\pi^+$ momenta in the
di-pion rest frame ($\theta$). The relations between ($E_\gamma^*$, $\theta$)
and the $z_i$ are:
$$
z_3={E_\gamma^* \over m_K}, \qquad\qquad z_\pm={E_\gamma^* \over 2 m_K}
\big(1\mp \beta \cos(\theta)\big), \eqno(B.1)
$$
where $\beta=\sqrt{1-4m^2_\pi/(m_K^2-2m_KE_\gamma^*)}$.
The kinematical limits on $E_\gamma^*$ and $\theta$ are given by:
$$
0<E_\gamma^* < { m_K^2 -4m^2_\pi \over 2m_K }, \qquad\qquad
-1 \le \cos(\theta) \le 1. \eqno(B.2)
$$
Finally the differential rate in terms of these variables is:
$$
{\partial^2\Gamma\over{\partial E_\gamma^* \partial \cos(\theta)}} =
{E_\gamma^*\beta \over 2 m^2_K} {\partial^2\Gamma\over
{\partial z_1\partial z_2}}
= \Big[ |E|^2 +|B|^2 \Big]  {E_\gamma^{*3}\beta^3
 \over 512 \pi^3 m_K^3 }
 \left(1- {2 E_\gamma^*\over m_K}\right)\sin^2(\theta) .
\eqno(B.3)
$$

In $\Koppg$ decays the situation is different, it is more useful to study
the differential rate as a function of [\abrams]: i) the charged pion kinetic
energy in the kaon rest frame ($T_c^*$), ii) the adimensional variable
$W^2=(qp_{_K})(qp_\pm)/(m_{\pi^+}^2m_K^2)$. These variables are related to the
$z_i$ by:
$$
z_\pm={1\over 2m_K^2}\left[m_K^2 +m_{\pi^+}^2-m_{\pi^0}^2-2m_Km_{\pi^+}
-2m_KT_c^*\right], \qquad
z_3z_\pm= {m_{\pi^+}^2\over m_K^2}W^2. \eqno(B.4)
$$
While for $T_c^*$ it is easy to write the limits:
$$
0<T_c^* < { (m_K -m_{\pi^+})^2 -m_{\pi^0}^2 \over 2m_K}, \eqno(B.5)
$$
for $W$ the expressions are cumbersome and we refer to the figure 3.3
of Ref.[\Report].
The advantage in using these variables lies in the fact that, through
the $W^2$-dependence, it is easier to disentangle the different contributions
of
inner bremsstrahlung, direct emission and  interference [\Good]:
$$
{\partial^2\Gamma\over{\partial T_c^* \partial W^2 }} =
\quad {\partial^2\Gamma_{IB} \over{\partial T_c^* \partial W^2 }}
\left\lbrace 1+ {m_{\pi^+}^2\over m_K} 2 \Real\left({E_{DE} \over e A}\right)
 W^2
 +{m_{\pi^+}^4\over m_K^2}\left(\left|{E_{DE} \over e A}\right|^2
+ \left|{M_{DE} \over e A}\right|^2 \right) W^4 \right\rbrace,
\eqno(B.6)
$$
where $A=A(\Kopp)$.

\vfill\eject

\vskip  1. true cm
\centerline {\bf References }
\vskip .8 true cm

\item{[\Sol]} V.V. Solov'ev and M.V. Terent'ev, JETP Lett.,
 {\bf{2}} (1965) 336. \par
  T.D. Lee and C.S. Wu, Annu. Rev. Nucl. Sci. {\bf{16}} (1966) 471;
 {\bf{16}} (1966) 511. \par
  A.D. Dolgov and L.A. Ponomarev, Soviet Phys. JNP {\bf{4}} (1967) 262. \par
  L. Sehgal and L. Wolfenstein, Phys. Rev. {\bf{162}} (1967) 1362. \par
  G. Costa and P.K. Kabir, Nuovo Cimento {\bf{A 51}} (1967) 564. \par
  M. McGuigan and A.I. Sanda, Phys. Rev.  {\bf{D 36}} (1987) 1413. \par
  Y.C.R. Lin and G. Valencia, Phys. Rev.  {\bf{D 37}} (1988) 143.  \par
\item{[\Report]} G. D'Ambrosio, M. Miragliuolo and P. Santorelli, \lq \lq
  Radiative non-leptonic kaon decays", appeared in \lq \lq
 {\it The DA$\Phi$NE Physics Handbook}\rq\rq, Eds. L. Maiani, G. Pancheri and
  N. Paver (1992).
\item{[\ENPa]}G. Ecker, H. Neufeld and A. Pich, Phys. Lett. {\bf B {278}}
  (1992) 337.
\item{[\ENPb]}G. Ecker, H. Neufeld and A. Pich, Nucl. Phys. {\bf B 314} (1994)
  321.
\item{[\Low]}  F.E. Low, Phys. Rev. {\bf 110} (1958) 974.
\item{[\WG]} S. Weinberg, Physica {\bf 96 A} (1979) 327. \par
  A. Manohar and H. Georgi, Nucl. Phys. {\bf B 234} (1984) 189. \par
  H. Georgi, {\it \lq\lq Weak Interactions and Modern Particle Theory"},
  Benjamin/Commings, Menlo Park 1984.
\item{[\GLa]} J. Gasser and H. Leutwyler, Ann. Phys. {\bf 158} (1984) 142;
  Nucl. Phys. {\bf B 250} (1985) 465.
\item{[\Ramberg]} E.J. Ramberg et al., Phys. Rev. Lett. {\bf 70} (1993) 2525.
\item{[\Seghal]} L.M. Seghal, Phys. Rev. {\bf D38} (1988) 808; Phys. Rev. {\bf
  D41} (1990) 1611.
\item{[\EPRc]} G. Ecker, A. Pich and E. de Rafael, Phys. Lett. {\bf B 237}
  (1990) 481, and references therein.
\item{[\DMS]} G. D'Ambrosio, M. Miragliuolo and F. Sannino,
Z. Physik {\bf C 59} (1993) 451.
\item{[\condon]} J.J. De Swart, Reviews of Modern Physics, {\bf 35} (1963)
916
 \item{[\EGPR]} G. Ecker, J. Gasser, A. Pich and E. de Rafael Nucl. Phys.
{\bf
  B 321}, (1989) 311. \par
  G. Ecker, H. Leutwyler, J. Gasser, A. Pich and E. de Rafael Phys. Lett. {\bf
  B 223}, (1989) 425. \par
  J.F. Donoghue, C. Ramirez and G. Valencia, Phys. Rev. {\bf D 39} (1989)
  1947. \par
  M. Praszalowicz and G. Valencia, Nucl. Phys. {\bf B 341} (1990) 27.
\item{[\BEG]}J. Bijnens, G. Ecker and J. Gasser,
  \lq\lq Introduction to chiral symmetry", appeared in \lq \lq
  {\it The DA$\Phi$NE Physics Handbook}\rq\rq, Eds. L. Maiani, G. Pancheri and
  N. Paver (1992), and references therein.
\item{[\Eckerb]}  G. Ecker, J. Kambor and D. Wyler, Nucl. Phys.
{\bf B 394} (1993) 101.
\item{[\Dib]}  C.O. Dib and R.D. Peccei, Phys. Lett. {\bf B 249} (1990) 325.
\item{[\Seghalc]} P. Heilinger and L.M. Seghal, Phys. Lett.
  {\bf B 307} (1993) 182.
\item{[\data]} Review of Particle Properties, Phys. Rev {\bf D 45} (1992).
\item{[\ochs ]} W. Ochs, \lq \lq Results On Low Energy $\pi \pi$ Scattering",
 Max-Plank Institute preprint MPI-Ph/Ph 91-35 (June 1991),
published in $\pi N$ Newsletters {\bf 3} (1991) 25.
\item{[\GM]} J. Gasser and U.G. Meissner, Phys. Lett. {\bf B 258} (1991)
 219.
\item{[\Rambergb]} E.J. Ramberg, private communication reported in Ref.[\ENPb].
\item{[\Esposito]}
  G. Ecker, Geometrical aspects of the non-leptonic
  weak interactions of
  mesons, in Proc. IX Int. Conf. on the {\it Problems of Quantum Field
  Theory}, M.K. Volkov (ed.), Dubna, 1990. \par
  G. Esposito-Farese, Z. Phys. {\bf C 50} (1991) 255.
\item{[\KMWnucl]} J. Kambor, J. Missimer and D. Wyler, Nucl. Phys. {\bf B 346}
  (1990) 17.
\item{[\Wolf1]}B. Winstein and L. Wolfenstein, Rev. Mod Phys. {\bf 65} (1993)
  1113.
\item{[\Taureg]} H. Taureg et al., Phys. Lett. {\bf B 65} (1976) 92.
\item{[\abrams]} R.J. Abrams et al., Phys. Rev. Lett. {\bf 29} (1972) 1118.
\item{[\Funck]} R. Funck and J. Kambor, Nucl. Phys. {\bf B 396} (1993) 53.
\item{[\EPRa]} G. Ecker, A. Pich and E. de Rafael, Nucl. Phys. {\bf B 303}
  (1988) 665.
\item{[\Good]} J.D. Good, Phys. Rev. {\bf 113} (1959) 352. \par
  N. Christ, Phys. Rev. {\bf 159} (1967) 1292.

\vfill\eject

\vskip  1. true cm
\centerline {\bf Tables and Figures}

\vskip  2. true cm
\centerline {\bf Table 1}
\vskip .5 true cm
Coefficients of the Taylor expansion of the functions
$h_{ij}$, for $ij=\pi\pi$, $\pi K$ and $K \eta$.
\vskip .8 true cm
$$
\matrix{ & h_{ij}^0  & h_{ij}^1  & h_{ij}^2  \cr    & & & \cr
 \pi\pi \qquad  & +.36-.15i & -.30+.22i & +.37-.42i \cr
 \pi K  \qquad  & -.12      & -.045     &<10^{-2}   \cr
 K \eta \qquad  & -.04      & -.005     &<10^{-2}   \cr}
$$

\vskip  2. true cm
\centerline {\bf Table 2}
\vskip .5 true cm
Inner bremsstrahlung and direct emission contributions
to the branching ratios of $\Ksppg$ decay, for different
values of the cut in the photon energy.  Different
values of the $o(p^4)$ counterterms are chosen.
\vskip .8 true cm
$$
\matrix{ & E_\gamma^* > 20 MeV & E_\gamma^* > 50 MeV  & E_\gamma^* > 100 MeV
\cr
 & & & \cr
 10^3\cdot IB\hfill & \hfill 4.80 \hskip15.pt & \hfill 1.75\hskip15.pt &
\hfill .31 \hskip 15.pt \cr
 10^6\cdot DE\ ( k_f=0) \hfill & \hfill -6.2 \qquad  & \hfill -5.0 \qquad &
\hfill -2.0 \qquad \cr
 10^6\cdot DE \ (k_f=+.5) \hfill & \hfill  -10.5 \qquad & \hfill -8.3\qquad
 & \hfill -3.3\qquad \cr
 10^6\cdot DE\ ( k_f=+1)\hfill & \hfill -14.8\qquad & \hfill -11.7\qquad
 & \hfill -4.7\qquad \cr
 10^6\cdot DE\  ( k_f=-.5)\hfill & \hfill -1.9\qquad & \hfill -1.6\qquad
 & \hfill -.6\qquad \cr
 10^6\cdot DE\ ( k_f=-1)\hfill & \hfill +2.4\qquad & \hfill +1.8\qquad
 & \hfill +.7\qquad \cr}
$$

\vskip  2. true cm
\centerline {\bf Figure 1}
\vskip .5 true cm
$o(p^4)$ loop diagrams for $\Kppg$ direct emission amplitudes.
The symbols $\circ$ and $\bullet$ represent the strong and the weak vertex,
respectively. The photon has to be attached to any charged line or to any
vertex.

\bye